\documentclass[showpacs,twocolumn,prl,amssymb,amsmath,nobibnotes,aps]{revtex4}
\usepackage{bm}

\newcommand{\xv}{{\bf x}}
\newcommand{\rv}{{\bf r}}
\newcommand{\qv}{{\bf q}}
\newcommand{\hR}{\hat{R}}
\newcommand{\inftyP}{{(\infty)}}
\newcommand{\D}{{(\cal{D})}}
\newcommand{\N}{{(\cal{N})}}
\newcommand{\ax}{{(a)}}

\newcommand{\be}{\begin{equation}}
\newcommand{\ee}{\end{equation}}
\newcommand{\bea}{\begin{eqnarray}}
\newcommand{\eea}{\end{eqnarray}}
\newcommand{\oh}{\frac{1}{2}}

\def\rf#1{(\ref{#1})}
\def\rfs#1{Eq.~\rf{#1}}
\usepackage[dvips]{graphicx}
\begin{document}

\title{Liquid crystal cells with ``dirty'' substrates} 
\author{Leo Radzihovsky} 
\author{Quan Zhang}
\affiliation{Department of Physics, University of Colorado,
   Boulder, CO 80309, USA}
\date{\today}

\begin{abstract}
  We explore liquid crystal order in a cell with a ``dirty'' substrate
  imposing a random surface pinning. Modeling such systems by a
  random-field xy model with {\em surface} heterogeneity, we find that
  orientational order in the three-dimensional system is marginally
  unstable to such surface pinning. We compute the Larkin length
  scale, and the corresponding surface and bulk distortions. On longer
  scales we calculate correlation functions using the functional
  renormalization group and matching methods, finding a universal
  logarithmic and double-logarithmic roughness in two and three
  dimensions, respectively. For a finite thickness cell, we explore
  the interplay of homogeneous-heterogeneous substrate pair and detail
  crossovers as a function of disorder strength and cell thickness.
\end{abstract}

\maketitle 

\begin{figure}[b]
\includegraphics[height=5 cm]{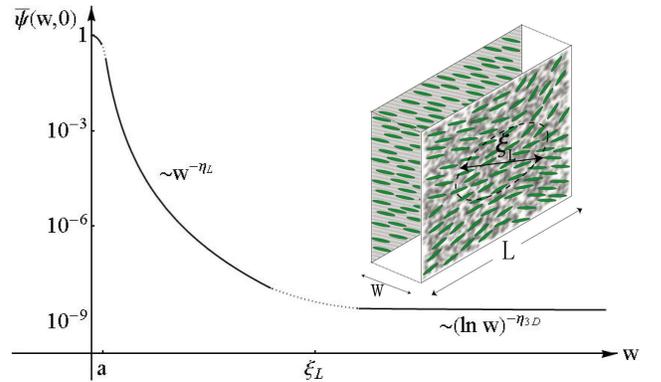}\\      
\caption{Average orientational order parameter $\overline{\psi}(w,0)$
  (controlling light transmission through a liquid-crystal cell) at
  the random pinning substrate of a 3D Dirichlet cell of
  thickness $w$, illustrated in the inset.}
\label{LCcell}
\end{figure}

Over the past several decades there has been considerable progress in understanding the
phenomenology of ordered states subject to random heterogeneities,
generically present in real materials \cite{FisherPhysicsToday}.
While most of the focus has justifiably been on the {\em bulk} heterogeneity, there are
many realizations in which random pinning is confined to a {\em surface} of
the sample. A technologically relevant example, illustrated as the inset of
Fig.\ref{LCcell}, is that of a liquid-crystal cell (e.g., of a laptop
display), where a ``dirty'' substrate imposes random pinning, that
competes with liquid-crystal ordering. Manifestations of heterogeneous
surface anchoring include common schlieren textures \cite{schlieren_texture} and multistability
effects\cite{Aryasova} observed in nematic cells, as well as long-scale smectic layer
distortions in smectic cells\cite{ClarkSmC}.

In this Letter we study such {\em surface} randomly-pinned systems
modeling them via a $d$-dimensional xy model with the heterogeneity
confined to a $(d-1)$-dimensional surface.  One might a priori expect
surface pinning effects to be vanishingly weak compared to the
ordering tendency of the homogeneous bulk, and thus the xy order to be
stable to weak surface disorder in {\em any} dimension. 

Our finding contrasts sharply with this intuition. Namely, our key
qualitative observation is that the xy order on the surface of such
$d$-dimensional system with $d\leq d_{lc}=3$ is always destabilized by
arbitrary weak random surface pinning\cite{FeldmanPRL}. 

Our finding can be understood from a generalization of the bulk
Imry-Ma argument\cite{Larkin,ImryMa} to the surface pinning
problem. For an ordered region of size $L$, the interaction with the surface random field
can lower the energy by $E_{pin}\sim
V_p\sqrt{N}_p\sim\Delta_f^{1/2}L^{(d-1)/2}$, where $V_p$ is a typical
pinning strength with zero mean and variance $\Delta_f\approx
V_p^2/\xi_0^{d-1}$ ($\xi_0$ is the pinning correlation length) and
$N_p$ is the number of surface pinning sites. Since surface
distortions on scale $L$ extend a distance $L$ into the bulk, the
corresponding elastic energy cost scales as $E_e\sim K L^{d-2}$, where
$K$ is the elastic stiffness. By comparing these energies it is clear that 
for $d<3$, on sufficiently long scales, $L >\xi_L\sim
(K^2/\Delta_f)^{(1/(3-d))}$ the surface heterogeneity dominates
over the elastic energy, and thus on these long scales always destroys
long-range xy order for arbitrarily weak surface
pinning\cite{FeldmanPRL}. A more detailed analysis extends the argument to 3D. 
The {\em surface} Larkin length scale\cite{Larkin}, beyond which the
orientational order on the $T=0$ random surface ($z=0$) is destroyed,
is given by
\begin{equation}
\xi_L\approx a e^{c K^2/\Delta_f},\ \ \ \ \ \ \mbox{for $d = 3$},
\label{xiL}
\end{equation}
where $c=8\pi^3$ is a nonuniversal constant, and $a$ is a microscopic 
cutoff of order of a few nanometers in the context of liquid crystals, 
set by the molecular size.

Although these (heterogeneous) substrate-induced distortions from
short scales $x<\xi_L$ only penetrate a distance $\xi_L$ into the
bulk, decaying exponentially beyond this scale, the bulk ordered state
never recovers. That is, strikingly, we find that logarithmically
rough distortions, persist an arbitrary long distance $z$ from the
dirty substrate, as summarized by
$C(\xv,z,z')=\overline{\langle(\phi(\xv,z)-\phi(0,z'))^2\rangle}$,
with (for $x\gg\xi_L$ in 2D)
\begin{eqnarray}
\hspace{-0.5cm} C^\inftyP_{2D}(\xv,z,z)&\approx&b_2 e^{-2z/\xi_L} + 
\frac{\pi^2}{9}\ln\left[1+\frac{x^2}{(2z+\xi_L)^2}\right], 
\label{C2d}
\end{eqnarray}
($b_2$ is a weak function of $2z/\xi_L$), and in 3D given by a double-logarithmic asymptotics \rf{C3d}.  

With an eye to liquid-crystal cell
applications\cite{Aryasova,ClarkSmC}, we extend our analysis to a
nematic cell of finite thickness $w$, with a heterogeneous front
substrate, and a back substrate with a homogeneous Dirichlet $\D$ or
Neumann $\N$ boundary condition, that can be imposed by various substrate
treatments, for example, surface polymer coating and rubbing. As expected, for the
$\cal{D}$-cell long-range orientational order is stable to weak
pinning, but with a value of {\em surface} orientational order
parameter $\overline{\psi}$ that exhibits a strong crossover as a
function of $w/\xi_L$ as shown in Fig.~(\ref{LCcell}), that in 3D is
given by
\begin{eqnarray}
\overline{\psi}\approx\left\{\begin{array}{ll} 
\left(\frac{a}{w}\right)^{\eta_L},& \mbox{thin cell, $a\ll w\ll\xi_L$},\\
e^{-\alpha}
\left[\frac{\ln(\xi_L/a)}{\ln(w/a)}\right]^{\eta_{3D}},& \mbox{thick cell, $w\gg\xi_L$},
\end{array}\right.,\ \ \ 
\label{psi}
\end{eqnarray}
where $\eta_{3D}=\pi^2/18$ is a universal exponent and $\alpha=2\pi^2$,
$\eta_L=2\pi^2/\ln(\xi_L/a)$ are nonuniversal constants. In contrast,
the long-range orientational order is unstable for the $\cal{N}$ cell,
with the ratio $w/\xi_L$ determining the rate of decay with the
transverse cell size, $L$.  


We now outline the analysis that leads to above results.  As a ``toy''
model of a nematic liquid-crystal cell\cite{toyComment} with a dirty
substrate and thickness $w$, we employ a $d$-dimensional {\em surface}
random-field xy model characterized by a Hamiltonian
\begin{eqnarray}
\hspace{-0.2cm}
H=\int_{-\infty}^{\infty}d^{d-1}x \int_0^wdz
\left[\frac{K}{2}(\nabla \phi(\rv))^2
-V[\phi(\rv),\xv]\delta(z)\right].\ \
\label{Hbulk}
\end{eqnarray}
In the equation above $\phi(\rv)$ is the xy-field distortion (azimuthal nematic
director angle in the plane of the substrate) at a point
$\rv\equiv(\xv,z)$, the random surface pinning potential
$V[\phi(\xv,z),\xv]\delta(z)$, characterized by a Gaussian
distribution with a variance $\overline{V(\phi,\xv)V(\phi',\xv')}
=R(\phi-\phi')\delta^{d-1}(\xv-\xv')$, is confined to a front
substrate at $z=0$, and at the back homogeneous substrate at $z=w$ we
impose either a Dirichlet [$\phi(\xv,z)|_{z=w}=0$] or a Neumann
[$\partial_z\phi(\xv,z)|_{z=w}=0$] boundary condition for the
$\cal{D}$ and $\cal{N}$ cells, respectively.  The long scale behavior
of the periodic (period $2\pi$) variance function $R(\phi)$
characterizes low-temperature properties of the system.

Because the random pinning potential in \rf{Hbulk} is confined to the
front substrate at $z=0$, no nonlinearities appear in the bulk
($0<z<w$) of the cell. Consequently, as in other boundary
problems\cite{tiltLR}, it is convenient to focus on the
random substrate and exactly
eliminate the bulk harmonic degrees of freedom $\phi(\xv,z)$ in favor
of the random substrate field
$\phi_0(\xv)\equiv\phi(\xv,z=0)$.\cite{tiltLR} This can be done via a
constrained path integral by integrating out $\phi(\xv,z)$ with a
constraint $\phi(\xv,z=0)=\phi_0(\xv)$, thereby obtaining an effective
$(d-1)$-dimensional Hamiltonian for $\phi_0(\xv)$
\cite{tiltLR}. Equivalently (for $T=0$ properties), we can eliminate
$\phi(\xv,z)$ by solving the bulk Euler-Lagrange equation
$\nabla^2\phi(\rv)=0$. To this end, we Fourier transform
$\phi(\xv,z)$ over $\xv$, obtaining an equation for
$\phi(\qv,z)=\int d^{d-1}x\phi(\xv,z)e^{-i\qv\cdot\xv}$,
$\partial_z^2\phi(\qv,z)-q^2\phi(\qv,z)=0$, whose solutions for
the boundary conditions of interest are obtained by elementary methods
summarized by $\phi^{(a)}(\qv,z)=\phi_0(\qv)\varphi^{(a)}(q,z)$, with
mode-functions $\varphi^{(a)}(q,z)$ given by: $e^{-q z}$
($a=\infty$), $\sinh{\left[q(w-z)\right]}/\sinh{(q w)}$
($a=\cal{D}$), $\cosh{\left[q(w-z)\right]}/\cosh{(q w)}$
($a=\cal{N}$).

Substituting these into \rf{Hbulk} we obtain a $(d-1)$-dimensional
(surface) Hamiltonian
$H_s=\int\frac{d^{d-1}q}{(2\pi)^{d-1}}\oh\Gamma_q^{(a)}|\phi_0(\qv)|^2
-\int d^{d-1}x V[\phi_0(\xv),\xv]$ for the field $\phi_0(\xv)$ confined to the
random substrate at $z=0$, with elastic kernels given by
\begin{eqnarray}
\Gamma_q^{(\infty)}
&=& K q,\hspace{1.5cm}w\rightarrow\infty,\label{GammaI}\\
\Gamma_q^{\D}
&=& K q \coth(q w),
\ \mbox{Dirichlet}\label{GammaD}\\
\Gamma_q^{\N}
&=& K q \tanh(q w),
\ \mbox{Neumann}\label{GammaN}.
\end{eqnarray}
The finite thickness $\Gamma_q^{\D}$ and $\Gamma_q^{\N}$ kernels reduce to $K
q$ for $w\rightarrow\infty$, 
as required. The $q$ nonanalyticity and
long wavelength stiffening (compared to the bulk $K q^2$ elasticity)
of the latter arises due to a mediation of surface distortions by
long-range deformations in the bulk of the cell. In the opposite limit
of a thin cell and long scales, as expected the Dirichlet kernel
reduces to a ``massive'' one, $K/w$, and the Neumann kernel simplifies
to $K w q^2$ of an ordinary surface (without a contact with the bulk)
xy model.

For convenience we use the standard replica ``trick''\cite{Anderson} to work
with a translationally invariant field theory, at the expense of introducing
$n$ replica fields (with the $n\rightarrow 0$ limit taken at the end of the calculation).
The disorder-averaged free energy is given by ${\overline
  F}=-T\overline{\ln Z}=-T\lim_{n\rightarrow 0}(\overline{Z^n}-1)/n$,
with
$\overline{Z^n}=\int[d\phi_0^\alpha]e^{-H_s^{(r)}[\phi_0^\alpha]/T}$,
where the effective translation invariant replicated Hamiltonian
$H_s^{(r)}[\phi_0^\alpha]$ is given by
\begin{eqnarray}
  H_s^{(r)}&=&\sum_{\alpha}^n\int\frac{d^{d-1}q}{(2\pi)^{d-1}}
  \oh\Gamma_q^{(a)}|\phi_0^\alpha(\qv)|^2\\
  &&-\frac{1}{2T}\sum_{\alpha,\beta}^n\int d^{d-1}x R[\phi_0^\alpha(\xv)-\phi_0^\beta(\xv)].\nonumber
\label{Hsr}
\end{eqnarray}
The advantage of this dimensional reduction is that formally the problem
becomes quite similar to that of the extensively studied bulk random
pinning\cite{DSFisherFRG,Nattermann,GiamarchiLedoussal} in one lower
dimension and with a modified elasticity encoded in $\Gamma_q^{(a)}$,
\rf{GammaI}-\rf{GammaN}.

The importance of surface pinning can be assessed by computing
distortions $\overline{\langle\phi^2\rangle}$ (dominated by the 
zero-temperature distortions) within the Larkin approximation\cite{Larkin}, 
which accounts to a random torque (linear) approximation,
$\tau(\xv)=\partial_{\phi_0}V[\phi_0(\xv),\xv]\bigg|_{\phi_0=0}$ to
the random potential $V(\phi_0,\xv)$, with inherited Gaussian
statistics and variance $\overline{\tau(\xv)\tau(\xv')}
=-R''(0)\delta^{d-1}(\xv-\xv')\equiv\Delta_f\delta^{d-1}(\xv-\xv')$.

Simple analysis gives $\overline{\langle\phi_0^2(\xv)\rangle}^{(a)}
\approx\int_\qv \frac{\Delta_f}{\left[\Gamma^{(a)}_q\right]^2}$, which we find
to diverge with surface extent $L$ for an infinitely thick cell,
$w\rightarrow\infty$ for $d\leq d_{lc}=3$, and for an $\cal{N}$ cell
of thickness $w$ for $d\leq d^{\N}_{lc}=5$. The latter is consistent
with the $d_{lc}^{bulk}=4$\cite{Larkin,ImryMa} for a $(d-1)$-dimensional
random-field xy model, to which a thin $\cal{N}$ cell reduces. In
contrast, for a $\cal{D}$ cell these orientational root-mean-squared
(rms) fluctuations are finite, but (as expected) grow with cell thickness $w$.

In a standard way we identify the substrate extent $L_*$ at which these rms fluctuations
grow to order $2\pi$ with the Larkin length scale
$\xi_L^{(a)}$\cite{Larkin}. For an infinite cell thickness we find
$\xi_L$ given in the introduction by \rfs{xiL}.
For a finite thickness $\cal{D}$ cell we find
\begin{eqnarray}
\xi_L^{(\cal{D})}\approx\left\{\begin{array}{ll}
\xi_L,& \xi_L\ll w,\\
\frac{c_d w^{\nu_d+1}}{(\xi_L^{c}-\xi_L)^{\nu_d}},& 
\xi_L\approx\xi_L^c,
\end{array}\right.
\label{xiD}
\end{eqnarray}
with $\xi_L^{c}=a_d w$ the ``critical'' Larkin length, $c_2=1, 
a_2\approx 1.71, \nu_2=1$, and $c_3\approx 0.79, a_3\approx 1.23, 
\nu_3=1/2$. Thus for a cell thicker than the Larkin length, $\xi_L$, 
the back $\cal{D}$
substrate has weak influence on the range of the xy order, dominated
by the random front substrate. However, for a thin cell, such that
$\xi_L$ spans the cell thickness, the $\cal{D}$ substrate effectively
enforces the xy-order alignment across cell, suppressing
$\phi_0^{rms}$ below $2\pi$ and thereby driving the cell Larkin
scale, $\xi_L^\D$ to diverge.

However, this divergence is {\em not} an indication of a sharp
transition. Rather, (as we will see below) it is a signal of a crossover 
from a weakly xy-ordered state (at strong disorder and thick cell) for $\xi_L\ll w$
to a strongly xy-ordered state (at weak disorder and thin cell) for
$\xi_L\gg w$. In both limits the aligning $\cal{D}$ substrate
dominates over the random one, leading to a long-range xy order.  For
the $\cal{N}$ cell we instead find
\begin{eqnarray}
\xi_L^\N\sim\left\{\begin{array}{ll}
\xi_L,& w\gg\xi_L,\\
w^{\gamma_d}\left\{\begin{array}{ll}
\sqrt{\ln(\xi_L/w)},&d=3,\\ 
(\xi_L)^{((3-d)/(5-d))},&d<3,\end{array}\right.&w\ll\xi_L, 
\end{array}\right.
\end{eqnarray}
with $\gamma_d =2/(5-d)$, lower limit corresponding to a thin $(d-1)$-
dimensional ``film'' pinned by $(d-1)$-dimensional ``bulk'' disorder.

On length scales longer than the crossover scale $\xi_L^{(a)}$,
$\phi_0$ distortions grow into a nonlinear regime, where random torque
model is clearly inadequate, and the effects of the random potential
$V[\phi_0(\xv),\xv]$ must be treated nonperturbatively. As with bulk 
disorder problems, this can be
done systematically using a functional renormalization group (FRG)
analysis\cite{DSFisherFRG,GiamarchiLedoussal} in an expansion in
$\epsilon=d_{lc}-d=3-d$.

To this end we employ the standard momentum-shell
FRG\cite{DSFisherFRG,GiamarchiLedoussal,RZunpublished}, integrating out
perturbatively (to one-loop) in
$R[\phi_{0}^\alpha(\xv)-\phi_{0}^\beta(\xv)]$ the short-scale
modes with support in an infinitesimal momentum shell $\Lambda/b < q <
\Lambda\equiv 1/a$, with $b=e^{\delta\ell}$. Taking $w(b) = b^{-1}
w$, for $T=0$ we obtain
\begin{eqnarray}
\partial_\ell\hR_{a}(\phi)=\epsilon^{(a)}(\ell)\hR_{a}(\phi)
+ \frac{1}{2} (\hR_{a}''(\phi))^2-\hR_{a}''(\phi)\hR_{a}''(0),\ \ \ 
\label{FRGflowRT0_DN}
\end{eqnarray}
with effective eigenvalues given by
$\epsilon^{(\cal{D},\cal{N})}(\ell)=\epsilon \mp\frac{4 \Lambda
  w(\ell)}{\sinh[2\Lambda w(\ell)]}$ and dimensionless disorder
variance functions defined by
$\hR_{\cal{D},\cal{N}}(\phi)\equiv\frac{C_{d-1}\Lambda^{d-3}}
{K^2\coth^{\pm 2}(\Lambda w)} R(\phi)$ for the $\cal{D}$ and
$\cal{N}$ choices of boundary conditions on the back substrate.
In Eq.(\ref{FRGflowRT0_DN}), the prime indicates a partial derivative with respect to
$\phi$, and $C_d=1/[\Gamma(d/2)2^{d-1}\pi^{d/2}]$.

For $w\rightarrow\infty$, aside from the Gaussian eigenvalue of
$\epsilon=3-d$ (rather than $4-d$), the flow of $\hR$ for our {\em
  surface} pinning problem reduces to that of the {\em bulk} pinning
problem
\cite{DSFisherFRG,GiamarchiLedoussal,ChitraGiamarchiDoussal,FeldmanPRL}.
In 3D the disorder is marginally irrelevant, vanishing according to
$\hR''(\phi,\ell)=\frac{1}{\ell}
\left[-\frac{1}{6}(\phi-\pi)^2+\frac{\pi^2}{18}\right]$, and flows to
a fixed point $\hR''_*(\phi)=\epsilon
\left[-\frac{1}{6}(\phi-\pi)^2+\frac{\pi^2}{18}\right]$ for
$d<3$ \cite{ChitraGiamarchiDoussal,FeldmanPRL}. In 2D $T$ is no longer
irrelevant leading to a qualitatively distinct behavior dominated by a
single harmonic of the random potential, which in turn leads to a
finite $T$ super-roughening transition\cite{RZunpublished,CO,TD}.  In the
opposite extreme of a microscopically thin cell, such that $w\ll a$,
the eigenvalues reduce to $\epsilon^\D(\ell)\approx 1-d$ and
$\epsilon^\N(\ell)\approx5-d$. These correspond to flows of a $(d-1)$-dimensional
{\em bulk} random-field xy model, with the $\cal{D}$ cell
effectively subjected to a uniform external field due to the back
substrate, dominating over the random pinning in the physical
dimensions ($d=2,3$).

For a more realistic case of a finite cell thickness $w$, there is a
crossover at scale $b_w^*\equiv w/a$ from a thick $d$ dimensional cell
at small $\ell$ such that $\Lambda w(\ell)\gg 1$ to an effective $(d-1)$-dimensional
 ``film'' for $\Lambda w(\ell)\ll 1$.  Another crossover
scale encoded in flow equations, \rfs{FRGflowRT0_DN}, is set by a
scale $b_L^*$ at which the nonlinear terms become comparable to the
linear ones, where the flow leaves the vicinity of the Gaussian fixed
point. It can be shown that this latter scale is simply set by the Larkin length, with
$b_L^*=\xi_L/a$.  As we will see blow, the nature of distortions
strongly depends on the relative size of these two crossover scales
and on the type of boundary condition on the uniform substrate. We
naturally designate the two cases, $w\ll \xi_L$ and $w\gg \xi_L$ as
thin and thick cells, respectively.

We can now utilize these FRG flows to compute the orientational
correlation function $C^\ax$ by establishing a relation for it at
different scales: $C^\ax[\qv,K,w,\hR_{a}]=e^{(d-1)\ell}C^\ax[\qv
e^{\ell},K(\ell),w(\ell),\hR_{a}(\ell)]$, where the prefactor comes from
the dimensional rescalings, keeping $T$ fixed. Choosing $\ell_*$ such
that $q e^{\ell_*}=\Lambda$, allows us to reexpress $\ell_*$ inside
$C^\ax$ in terms of $q$. This gives
$C^\ax[\qv,K,w,\hR_{a}]\approx-\left(\frac{\Lambda}{q}\right)^{d-1}
\frac{R''(0,\ell_*)}{\left[\Gamma_\Lambda^{(a)}(\ell_*)\right]^2}$,
which explicitly requires an analysis of the flow for specific
boundary conditions.

For an infinitely thick cell ($w\rightarrow\infty$) and $q<1/\xi_L$
this gives in $d<3$:
$C^\inftyP_*[\qv,K,\infty,\hR]\approx\frac{1}{q^{d-1}}
\frac{(3-d)\pi^2}{9C_{d-1}}$, and in 3D:
$C^\inftyP_*[\qv,K,\infty,\hR]\approx\frac{-1}{q^2\ln(q
  a)}\frac{\pi^2}{9C_2}$, from which real space correlations on
in-plane scales $x > \xi_L$ (inaccessible within Larkin approximation)
can be computed.  Fourier-transforming $C^\inftyP(\qv)$ we find
$C^\inftyP(\xv,z,z) \approx C^\inftyP_L(\xv,z)+C^\inftyP_*(\xv,z)$,
where $C^\inftyP_L(\xv,z)$ is the contribution from short scales,
$\xi^{-1}_L< q < a^{-1}$, where random torque model is valid, given by
$C_L^\inftyP(\xv,z,z)=
(3-d)8\pi^2\left(\frac{2z}{\xi_L}\right)^{3-d}\Gamma(d-3,2z/\xi_L,2z/a)
\approx b_d e^{-2z/\xi_L}$ with $\Gamma(p,z_1,z_2)=\int_{z_1}^{z_2}
t^{p-1} e^{-t} dt$ the generalized incomplete Gamma function.  The
second long-scale part, $C^\inftyP_*(\xv,z)$ is a universal
contribution, a Fourier transform of $C^\inftyP_*[\qv,K,\infty,\hR]$,
which when combined with $C_L^\inftyP$ at low $T$, in 2D gives
\rfs{C2d}. In 3D this matching procedure gives\cite{FeldmanPRL}
\begin{eqnarray}
C^\inftyP_{3D}(\xv,z,z)&\sim&
\frac{2\pi^2}{9}\left\{\begin{array}{ll}
\ln\big[\frac{\ln(x/a)}{\ln(\xi_L/a)}\big],&2z \ll \xi_L \ll x\\
\ln\big[\frac{\ln(x/a)}{\ln(2z/a)}\big],&\xi_L\ll 2z \ll x\\ 
\frac{x^2}{16z^2}\frac{1}{\ln{(2z/a)}},&\xi_L \ll x \ll 2z 
\end{array}\right..\nonumber\\
\label{C3d}
\end{eqnarray}

Another quantity of interest is the average orientational order
parameter, $\overline{\psi}(z)=\overline{\langle e^{i\phi}\rangle}
\approx e^{-\overline{\langle\phi^2\rangle}/2}$, where
somewhat crudely we approximated it by assuming Gaussian-correlated
$\phi(\rv)$.

From $C^\ax(\qv)$ for $w\rightarrow\infty$ and for arbitrarily thick
$\cal{N}$ cell we find that
$\phi_{rms}^2(L,z)=\overline{\langle\phi^2\rangle}$ grows without
bound with planar cell extent $L$. Thus for $a=\infty$ and $a=\cal{N}$
cells in $d\leq 3$ the orientational order parameters,
$\overline{\psi}^{(\infty,\cal{N})}$ vanish in the thermodynamic limit
for arbitrarily weak surface heterogeneity.

In contrast, in a $\cal{D}$ cell the growth of $\phi_{rms}^2(L,z)$ is
suppressed by the aligning homogeneous $\cal{D}$ substrate. Thus the
order is stable for an arbitrarily thick (but finite) cell, characterized by
$\overline{\psi}(w,z)$ that is a strong function of $w/\xi_L$,
computable via above matching analysis.  For a thin cell
($w\ll\xi_L$), $w(\ell)=e^{-\ell}w$ reaches the microscopic scale $a$
at $e^{\ell_w^*}=w/a$ and therefore $\epsilon^\D(\ell>\ell_w^*)\approx
\epsilon-2=1-d$ {\em before} $e^{\ell_L^*}=\xi_L/a$. Since beyond
$\ell_w^*$, $\epsilon^\D(\ell>\ell_w^*)<0$, pinning is irrelevant with
its growth cutoff at scale $e^{\ell_w^*}$. In this case, scales beyond
$\xi_L$ are not probed (the flow never leaves the vicinity of the
Gaussian fixed point) and $\phi_{rms}(w,z)$ can be accurately computed
within the random torque model, cut off by $w$. Thus, utilizing our
earlier definition of $\xi_L$, for a {\em thin} $\cal{D}$ cell in
$d<3$, we find $\phi_{rms}^2(w,0)
\approx4\pi^2\left(\frac{w}{\xi_L}\right)^{3-d}$.

For a thick $\cal{D}$ cell ($w\gg\xi_L$), the flow instead crosses
over to the vicinity of the nontrivial fixed point $R_*(\phi)$
(leaves the Gaussian fixed point) {\em before} it is cutoff by the
finite $w$. Thus on these intermediate scales (defined by $w/a\equiv
e^{\ell_w^*} > e^\ell > e^{\ell_L^*}\equiv\xi_L/a$) despite being
ultimately cutoff by $w$, $\phi_0$ distortions are large, requiring
a nonperturbative treatment of surface pinning. As for above
computation of $C^\inftyP(\xv)$, in this regime
$\overline{\psi}(w,0)=e^{-\phi_{rms}^2(w,0)/2}$ can be well
estimated by the matching FRG analysis. 

Thus, neglecting the subdominant contribution from scales longer than
$w$, approximating the $\phi_0$ correlator $C(\qv)$ by its
fixed-point value, valid for $w\gg \xi_L$ [such that the flow
approaches the vicinity of the nontrivial fixed point,
$R_*(\phi,\ell)$], and approximating $C(\qv)$ by its Gaussian fixed
point expression, valid for $\xi^{-1}_L<q<a^{-1}$, for a thick cell
$w\gg\xi_L$ and $d<3$, we find
$\overline{\langle\phi_0^2\rangle}\approx
\frac{\epsilon\pi^2}{9}\ln(w/\xi_L) + 4\pi^2$. Putting these
crossovers together for $d<3$ we find
\begin{eqnarray}
\overline{\psi}(w,0)\approx\left\{\begin{array}{ll}
e^{-2\pi^2(w/\xi_L)^{3-d}},& \mbox{thin cell, $w\ll\xi_L$},\\
e^{-2\pi^2}\left(\frac{\xi_L}{w}\right)^{\eta_d^*}, & \mbox{thick cell, $w\gg\xi_L$}.
\end{array}\right.
\end{eqnarray}

In 3D such matching analysis gives $\overline{\phi_0^2}
\approx4\pi^2\frac{\ln(w/a)}{\ln(\xi_L/a)}$ for a thin cell
($w\ll\xi_L$), and $\overline{\phi_0^2}
\approx\frac{\pi^2}{9}\ln\left[\frac{\ln(w/a)}{\ln(\xi_L/a)}\right] +
4\pi^2$ for a thick cell ($w\gg\xi_L$), leading to
$\overline{\psi}(w,0)$ in \rf{psi}.

To conclude, above we have studied stability of an ordered xy model to
random {\em surface} pinning and discussed these results in the
context of a nematic liquid-crystal cell with a ``dirty'' nonrubbed
substrate. We found that for a thick 3D cell, at long scales, the
nematic order is marginally unstable to such surface pinning, and
computed the extent of the orientational order in cells of finite
thickness with a second homogeneous substrate.  We expect these
predictions to be testable via polarizer-analyzer transmission and
confocal microscopies, and through birefringence response to an
in-plane electric or magnetic field.

We leave the challenging question of topological defects
proliferation, generalizations to other interesting (e.g.,
smectic\cite{ClarkSmC}) states, and glassy nonequilibrium relaxation
(memory effects, aging, etc., all expected in our system) to future studies.

We thank N. Clark and S. Todari for discussions and acknowledgement
support by the NSF DMR-0321848, MRSEC DMR-0213918 (LR, QZ), the
Berkeley Miller and the University of Colorado Faculty Fellowships
(LR).  L.R. thanks Berkeley Physics Department for its hospitality
during part of this work.

{\em Note Added:} While our Letter was under review, we learned of an
earlier seminal work\cite{FeldmanPRL}, with significant overlap with
our infinite cell thickness limit. Where overlap exists, our results
are in complete agreement.


\begin{thebibliography}{99}
\bibitem{FisherPhysicsToday} D. S. Fisher, G. M. Grinstein, A. Khurana, \textit{Theory of Random Magnets}, Phys. Today {\bf 41}, No.12, 56 (1988).
\bibitem{schlieren_texture} {\em Defects in Liquid Crystals: Computer Simulations, Theory and Experiments}, 
Edited by O. D. Lavrentovich, P. Pasini, C. Zannoni and S. Zumer (Erice, Sicily, Italy 2000). 
\bibitem{Aryasova} N. Aryasova et al., Mol. Cryst. Liq. Cryst., Vol. {\bf 412}, pp. 351 (2004).
\bibitem{ClarkSmC} C. D. Jones, N. A. Clark, Bull. Am. Phys. Soc. {\bf
    49}, 307 (2004).
\bibitem{FeldmanPRL} D. Feldman and V. Vinokur, Phys. Rev. Lett. {\bf 89}, 
227204 (2002).
\bibitem{Larkin} A. Larkin, Sov. Phys. JETP {\bf 31}, 784 (1970).
\bibitem{ImryMa} Y. Imry and S. K. Ma, Phys. Rev. Lett. {\bf 35}, 1399 (1975).
\bibitem{toyComment} A realistic model must include {\em nonplanar}
  director distortions, as well as point and line ($1/2$-disclinations) defects allowed by the headless nematic director.
\bibitem{tiltLR} L. Radzihovsky, Phys. Rev. B {\bf 73}, 104504 (2006).
\bibitem{Anderson} S. F. Edwards and P. W. Anderson, J. Phys. F {\bf 5}, 965 (1975).
\bibitem{Nattermann} T. Nattermann, Phys. Rev. Lett. {\bf 64}, 2454 (1990).
\bibitem{DSFisherFRG} D. S. Fisher,  Phys. Rev. B {\bf 31}, 7233 (1985).
\bibitem{GiamarchiLedoussal} T. Giamarchi, P. Le Doussal, Phys. Rev. B {\bf 52}, 1242 (1995).
\bibitem{RZunpublished} L. Radzihovsky and Q. Zhang, (to be published).
\bibitem{ChitraGiamarchiDoussal} R. Chitra et al., Phys. Rev. B {\bf 59}, 4058 (1999).
\bibitem{CO} J. L. Cardy, S. Ostlund, Phys. Rev. B {\bf 25}, 6899 (1982).
\bibitem{TD} J. Toner, D. P. DiVincenzo, Phys. Rev. B {\bf 41}, 632 (1990).

\end{thebibliography}
\end{document}